\documentclass[aps,prb,twocolumn,letterpaper,showpacs]{revtex4-1}
\usepackage[utf8]{inputenc}
\usepackage{graphicx,amsmath}
\usepackage{amssymb}
\usepackage{amsthm}
\usepackage[normalem]{ulem}

\usepackage{color}

\begin{document}
\title{Fluctuation driven topological Hund insulators}
\author{Jan Carl Budich$^{1}$, Bj\"orn Trauzettel$^{2}$, and Giorgio Sangiovanni$^{2}$}

\affiliation{$^1$Department of Physics, Stockholm University, Se-106 91 Stockholm, Sweden;\\
 $^2$Institute for Theoretical Physics and Astrophysics,
 University of W$\ddot{u}$rzburg, 97074 W$\ddot{u}$rzburg, Germany}
\date{\today}
\begin{abstract}
We investigate the role of electron-electron interaction in a two-band Hubbard model based on the Bernevig-Hughes-Zhang Hamiltonian exhibiting the quantum spin Hall (QSH) effect. By means of dynamical mean-field theory, we find that a system with topologically trivial non-interacting parameters can be driven into a QSH phase at finite interaction strength by virtue of local dynamical fluctuations. For very strong interaction, the system reenters a trivial insulating phase by going through a Mott transition. We obtain the phase diagram of our model by direct calculation of the bulk topological invariant of the interacting system in terms of its single particle Green's function.
\end{abstract}
\maketitle

\section{Introduction}
Triggered by the theoretical prediction \cite{KaneMele2005a,KaneMele2005b,BHZ2006} and experimental observation \cite{koenig2007} of the quantum spin Hall (QSH) state, topological properties of non-interacting band-structures have been a major focus of condensed matter physics in recent years \cite{KaneHasan,XLReview2010,VolovikReview,TSMReview}. An exhaustive enumeration of all possible topological band structures in the ten Altland-Zirnbauer symmetry classes \cite{AltlandZirnbauer} of insulators and mean field superconductors has been achieved using complementary approaches in Refs. \onlinecite{Schnyder2008,KitaevPeriodic,RyuLudwig} which concludes the non-interacting classification of these states. Since electron-electron interactions are to some extent present in every realistic system, the influence of electronic correlations on these topological bandstructures is among the key issues of this rapidly growing field \cite{MartinFakherReview}.

In the presence of interactions, the Hamiltonian is no longer a quadratic form in the field operators and hence a classification based on single particle Bloch-states is not possible. Instead, for interacting systems, the topological invariant can be formulated in terms of the single particle Green's function \cite{Redlich1984,So1985,Ishikawa1986,Ishikawa1987,VolovikQH3HE,Golterman1993,Volovik,QiTFT,TopologicalOrderParameter}. This was first proven \cite{Redlich1984,So1985,Ishikawa1986,Ishikawa1987,VolovikQH3HE,Golterman1993,Volovik} for the interacting generalization of the first Chern number characterizing the integer quantum Hall state \cite{TKNN1982,Kohmoto1985} and has more recently been generalized to symmetry-protected topological states \cite{QiTFT,TopologicalOrderParameter,WangTSC}. When switching on interactions adiabatically, the non-interacting invariant connects adiabatically to the mentioned Green's function invariant \cite{QiTFT,TopologicalOrderParameter}.

A natural question to ask is in what sense the Green's function topology of a system can change in the presence of interactions. On the one hand, interactions can renormalize the band parameters thus giving rise to a new effective non-interacting band structure. On the other hand, dynamical quantum fluctuations can change the $\omega$-dependence of the Green's function which has no non-interacting analogue: The Matsubara Green's function of a non-interacting system is always of the form $G(\omega,k)=(i\omega-h(k))^{-1}$. Deviations from this generic frequency dependence due to dynamical fluctuations have been shown to be capable of changing the value of the topological invariant by changing the pole structure of the inverse Green's function in $\omega$ \cite{Gurarie2011,FDWN2011,PE,localself}. Very recently, a significant simplification of the analytical form of the Green's function invariant has been achieved \cite{WangInversion,WangGeneralTOP,topham} which makes its practical calculation much more viable.

\section{Main results}
\begin{figure}[th]
\includegraphics[width=0.99\linewidth]{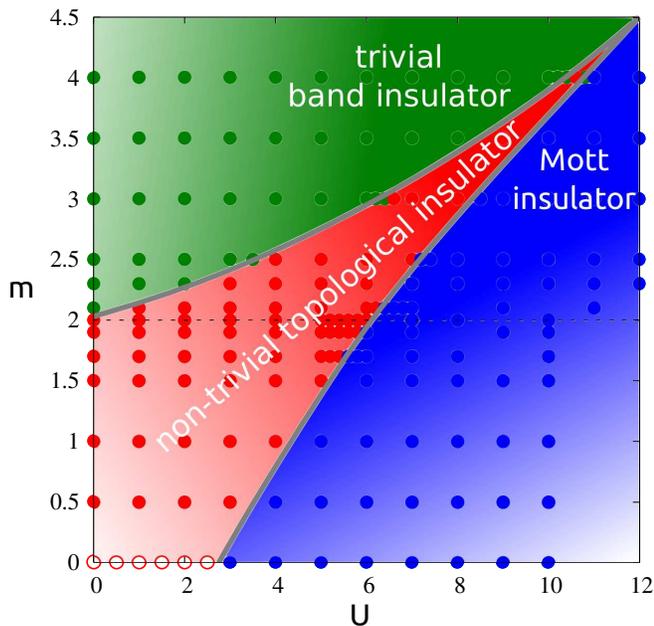}
\caption{Phase diagram of the BHZ model with $U$, $V\!=\!U-2J$ and $J\!=\!0.25U,~\lambda=0.3$. For $m\!>\!2$ the non-interacting model is topologically trivial (green) and for $m\!<\!2$ is in a topologically nontrivial QSH phase (red). At large values of $U$ the system is in a Mott phase (blue). The red empty circles at $m\!=\!0$ indicate that the solution is metallic. The grey lines denote transition regions. The ``topological Hund insulator'', i.e. the non-trivial interacting phase in red, can be obtained for small-to-moderate values of $U$, starting from a trivial phase with $m\!\gtrsim\!2$.}
\label{fig:phasediagram}
\end{figure}

In this work, we study the topological properties of an interacting two-band Hamiltonian based on the Bernevig-Hughes-Zhang (BHZ) model \cite{BHZ2006}. Besides its quantitative relevance for the low energy properties of HgTe quantum wells, for which it has been originally proposed, the BHZ model represents a generic minimal model for the QSH effect on a square lattice with two orbitals per site. 
Here, we hence consider it as the underlying non-interacting band structure of a two-orbital model, in combination with the most general on-site electron-electron interaction between two orbitals. We address the phase with no spontaneous symmetry breaking by means of dynamical mean field theory.
We show that starting from a topologically trivial band-insulator the system can be driven into a QSH phase upon increasing the interaction strength. For this to happen the interaction does not need to be particularly strong but it has to follow Hund's rule, i.e., favour the state with the highest total spin. For this reason we name this phase ``topological Hund insulator''. The relevance of high-spin many-body configurations is a natural tendency of multi-orbital models. This makes the trivial-to-nontrivial transition we find a very robust feature of interacting models for topological insulators.
We demonstrate that, if the band gap of the non-interacting system is sufficiently large, the dynamical quantum fluctuations are responsible for the topological transition, whereas the Hartree part of the self-energy stays in the trivial regime.
At larger interaction strengths a first-order Mott transition is found. The resulting strong-coupling phase is also associated with the low-frequency dependence of the Green's function so that its topology cannot be captured by static mean field theory. Our results are obtained by direct calculation of the Green's function homotopy in the presence of inversion symmetry as proposed in Ref. \onlinecite{WangInversion}.  The phase diagram is shown in Fig. \ref{fig:phasediagram}.

\section{Model and methods}
Previous literature on correlation effects in the QSH phase was mainly concerned with an on-site Hubbard model based on the Kane-Mele model for the QSH effect in graphene \cite{Rachel2010,Hohenadler2011,Zheng2011,MartinKM2012}. This honeycomb model has one orbital per spin for each carbon atom, i.e., for each site of the honeycomb lattice. In total this makes two states per spin per unit cell since each unit cell consists of two sites. However, as an on-site Hubbard model, the Kane-Mele model can be seen as a spinful single band model. The BHZ model in contrast takes into account one electron-like orbital (E) and one hole-like orbital (H) for each spin. Hence, a lattice version based on this model should be naturally considered as a two-band Hubbard model. We consider the Hamiltonian
\begin{align}
H= H_{\text{BHZ}}+H_{\text{int}},
\label{eqn:ham}
\end{align}
where $H_{\text{BHZ}}$ is the single particle Hamiltonian associated with the Bloch-Hamiltonian $H_{\text{BHZ}}(k)=\text{diag}\left(h(k),h^*(-k)\right)$, where the diagonal matrix structure is in spin space \cite{note}, $h(k)=\left(m-\cos (k_x)-\cos(k_y)\right)\sigma_z+\lambda \sin(k_x)\sigma_x+\lambda \sin(ky)\sigma_y$, and $\sigma_i$ are Pauli matrices in the E-H band pseudo spin space.
For the interacting part we take the most general two-orbital interaction parametrization $H_{\text{int}}=H_U+H_V+H_J$ \cite{GeorgesHundReview}.
$H_U=U\sum_i\left(n^{(E)}_{i,\uparrow}n^{(E)}_{i,\downarrow}+n^{(H)}_{i,\uparrow}n^{(H)}_{i,\downarrow}\right)$,
represents an intra-orbital Hubbard repulsion,
$H_V=V\sum_i\left(n^{(E)}_{i,\uparrow}n^{(H)}_{i,\downarrow}+n^{(H)}_{i,\uparrow}n^{(E)}_{i,\downarrow}\right)$,
and
$H_J=(V-J)\sum_i\left(n^{(H)}_{i,\uparrow}n^{(E)}_{i,\uparrow}+n^{(H)}_{i,\downarrow}n^{(E)}_{i,\downarrow}\right)$,
an inter-orbital term for electrons with opposite and parallel spin, respectively.
We also consider an $S_+S_-$-like contribution, namely
$H_S = -J \left(c^\dagger_{H,\downarrow} c^\dagger_{E,\uparrow} c^{\phantom{\dagger}}_{E,\downarrow} c^{\phantom{\dagger}}_{H,\uparrow} + c^\dagger_{H,\uparrow} c^\dagger_{E,\downarrow} c^{\phantom{\dagger}}_{E,\uparrow} c^{\phantom{\dagger}}_{H,\downarrow}\right)$
which makes the interaction fully SU(2)-symmetric.
In order to check to what extent our results depend on the details of the interaction, we consider three forms of $H_{\text int}$: $i$) $H_U+H_V+H_J$ with $V=U-2J$ and $J=0.25U$, $ii$) $H_U+H_V+H_J+H_S$ with the same values of $V$ and $J$ and $iii$) $H_U$ alone.
Hitherto, only the latter has been considered in the literature as interaction for the BHZ model (see Refs.\onlinecite{Yoshida2012,Tada2012,Medhi2012,WangBHZ}). Yet, the ``$U$-only'' interaction represents a very unreasonable choice for a multi-orbital Coulomb vertex, because the inter-orbital matrix elements are entirely neglegted. Case $i$) and $ii$) are therefore much closer to a realistic Hamiltonian for a realizable material.

We solve the model within dynamical mean field theory (DMFT). DMFT – for a review see [43] – corresponds to a mean-field approximation for the spatial degrees of freedom while preserving the local quantum dynamics of the interacting system. The self-energy $\Sigma(i\omega)$ depends therefore on the Matsubara frequencies only and the $k$-dependence of the Green’s function comes entirely from the one-particle Hamiltonian $H_{\text{BHZ}}$. This is a good approximation whenever the local correlations are the dominant ones (e.g. for high dimensionality, or at high temperatures, or far away from quantum critical points). With these limitations in mind, we conclude that DMFT represents a very powerful tool for the present study which focuses on the way the self-energy -- and hence the topology of the Green's function -- is modified by dynamical effects of the electron-electron interaction. DMFT allows us to go one step beyond static mean-field treatments which, by construction, cannot address the $\omega$-dependence of the self-energy and, contrary to DMFT, cannot yield dynamics-induced insulating states without long-range order.
As a DMFT-impurity solver we use the hybridization-expansion version of the continuous-time quantum Monte Carlo. The details of the code used for the numerical results can be found in Ref. [44]. The advantage of this flavour of impurity solver is that very low temperatures can be accessed (here we show results for $k_B T = 1/100$~ in units of the hopping parameter of $H_{\text{BHZ}}$, but we even went down to $1/400$) and the calculation is sign-problem free.

We consider the half-filled case, i.e. two electrons on two orbitals. The E band lays $2m$ above the H band, therefore, without interaction, the system is in the $(n_H,n_E)=(2,0)$ configuration. $U$ alone favors the $(1,1)$ configuration. On the other hand, the inter-orbital repulsion tends to restore $(2,0)$. The Hund coupling $J$ therefore plays the crucial role for favoring $(1,1)$ in the strong coupling region \cite{Werner,Sentef}.
The effective reduction of the energy difference between the orbitals is the important prerequisite to get the topological phase in the intermediate region. Indeed, were the bands effectively getting further and further apart from each other, the system would just forever stay in the trivial band-insulating $(2,0)$ configuration. This is for instance what would happen if we put $J\!=\!0$.

\begin{figure}[th]
\includegraphics[width=0.99\linewidth]{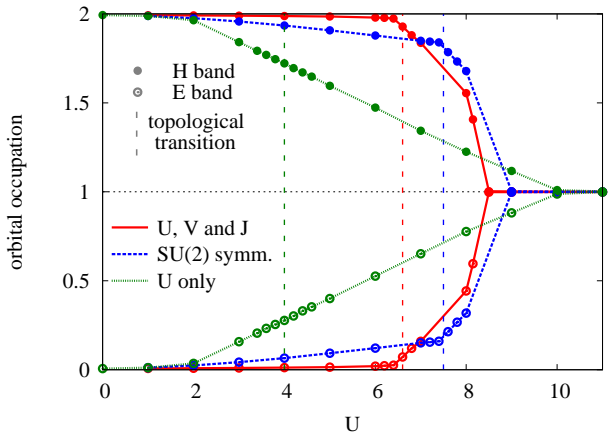}
\caption{(Color online) Orbital occupation as a function of the interaction strength for cases $i$) $H_U+H_V+H_J$ (red), $ii$) $H_U+H_V+H_J+H_{S}$ (blue) and $iii$) $H_U$ (green). All plots show calculations performed at $k_BT\!=\!1/100$, $m\!=\!3$ and $\lambda\!=\!0.3$. The vertical dashed lines mark the topological transition, while the Mott transition takes place when $(n_H,n_E)=(1,1)$.}
\label{fig:occ}
\end{figure}

\begin{figure}[th]
\includegraphics[width=0.99\linewidth]{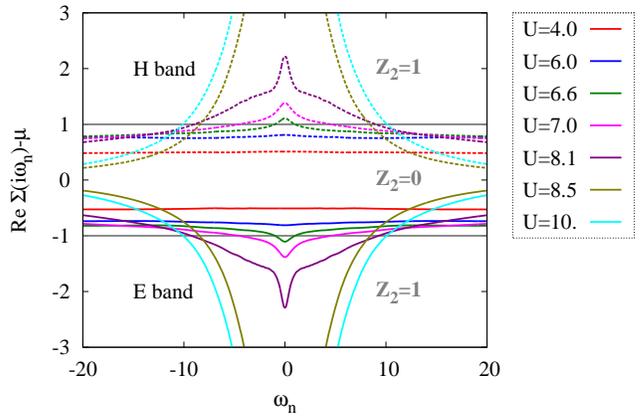}
\caption{(Color online) $\omega$-dependence of the real part of the self energy. Dashed lines denote the H bands. Solid lines denote the E bands. All plots show calculations performed at inverse temperature $\beta=100$, $J=U/4,~m=3,\lambda=0.3$.}
\label{fig:omegadep}
\end{figure}

In Fig. \ref{fig:occ}, we show the evolution of the orbital occupations for all three interaction forms.
We see how in all cases the system goes from the $(2,0)$ to the $(1,1)$ configuration upon increasing the interaction strength.
Case $iii$) gives an essentially linear decrease of the orbital polarization (confirming the oddness of such a choice of interaction) and, as a consequence, yields the topological transition in the weak-coupling regime where the frequency dependence of the self-energy is weak. As a consequence, the character of this transition is strongly static mean-field like. In contrast, cases $i$) and $ii$) are much richer and physically more interesting. Indeed, the topological transition is shifted to larger coupling, where the real part of the self-energy has a pronounced structure at low Matsubara frequencies. The dynamical fluctuations are therefore the driving force for the topological phase transition, as one can see in Fig. \ref{fig:omegadep} where the real part of the self-energy for case $i$) is shown for different values of $U$.

We now illustrate how we determine the topological properties from the low-frequency behavior of the real part of the DMFT self-energy $\Sigma(i\omega \! \rightarrow \! 0)$.
The explicit analytical form of the topological Green's function $\mathbb Z_2$-invariant characterizing the QSH state as originally proposed in Ref. \cite{QiTFT,TopologicalOrderParameter} is rather cumbersome to evaluate. This is because it employs a dimensional extension to the (4+1)D analogue of the quantum Hall effect \onlinecite{ZhangHu4DQH} in the framework of topological field theory and hence involves a five-fold frequency-momentum integration. However, in the presence of inversion symmetry a tremendous simplification of this complicated form has recently been achieved \cite{WangInversion}. Following also the more general analysis of Refs. \onlinecite{WangGeneralTOP,topham}, the recipe for the evaluation of the topological invariant in the presence of inversion symmetry that we employ here can be summarized as follows: First, one defines a fictitious non-interacting Hamiltonian \cite{WangGeneralTOP}
\begin{align}
\tilde H(k)=-G^{-1}(\omega=0,k)=H(k)+\Sigma(\omega \! \rightarrow \!0)
\label{eqn:topham}
\end{align}
which has been coined topological Hamiltonian \cite{topham}.
Except for values of the parameters exactly on the transition lines and for $m\!=\!0$ before the Mott phase, our solutions have insulating character and are characterized by a linearly vanishing Im $\Sigma$ and a finite extrapolation of Re $\Sigma(\omega \! \rightarrow \!0)$. The latter value becomes large in absolute value when entering the Mott phase, as shown in Fig. \ref{fig:omegadep}.

One can therefore proceed by calculating the $\mathbb Z_2$ invariant as proposed for the non-interacting case in the presence of inversion symmetry in Ref. \cite{FuInversion2007}. For the reader's convenience, we briefly review the procedure here. First, one calculates the eigenstates $\lvert \Gamma_i,\alpha\rangle$ of $\tilde H$ at the four time reversal invariant momenta (TRIM) $\Gamma_i,~i=1\ldots 4$, where $\alpha$ labels the occupied bands of $\tilde H$.
Due to inversion symmetry, these states can also be chosen as eigenstates of the parity operator $\mathcal P$ with eigenvalues $p_{i,\alpha}=\pm 1$. Next, using the phase convention $\sqrt{-1}=i$, one computes $\delta_{i,\alpha}=\sqrt{p_{i,\alpha}}$. Finally, the $\mathbb Z_2$ invariant $\nu$ can be expressed as \cite{FuInversion2007,WangInversion}
\begin{align}
(-1)^\nu=\prod_{i,\alpha}\delta_{i,\alpha}.
\label{eqn:z2}
\end{align}

\section{Phase diagram of the BHZ two-band Hubbard model}
We calculate the phase diagram of Fig. \ref{fig:phasediagram} using the interaction $i$) with $U,~ V=U-2J$ and $J\!=\!U/4$. The non-interacting model is in the QSH phase for $\vert m\rvert<2$ and in the trivial phase for $\lvert m\rvert>2$.
Fig. \ref{fig:phasediagram} shows cleary that for values of $2<m<4$, the system goes through three different phases as the interaction strength is increased: In the non-interacting limit, the system is topologically trivial in this regime, i.e., a normal band insulator (green region in Fig. \ref{fig:phasediagram}). At intermediate interaction strength, the system is driven into a QSH phase (red region) and in the strongly correlated limit the system is in a topologically trivial Mott phase (blue region). On the line separating the trivial (green) from the non-trivial (red) topological insulator, the system is a semi-metal. The Mott transition that follows is of first-order (for $m=3$, we observe an hysteresis region of about 0.1 width in $U$). An interesting open question is therefore what happens for $U>12$~ and $m>4.5$, after the two transition lines merge together. In a different system, namely strongly interacting Majorana modes in an array of Josephson junctions, a similar phase diagram has recently been predicted \cite{Has2012}.

An analysis at fixed $m\! = \!3$ of the evolution of the orbital occupation and of the self-energy shown, respectively, in Fig. \ref{fig:occ} and Fig. \ref{fig:omegadep}, illustrates well the nature of the two transitions.
As $U$ is increased from the non-interacting case the orbital polarization gets smaller (Fig. \ref{fig:occ}) therefore the bands start to effectively get closer to one another.
Since the interaction is inducing a pronounced frequency-structure in Re$\Sigma(i\omega)$, at $U \!=\! 6.6$ (green curves in Fig. \ref{fig:omegadep}), the zero-frequency values exceed in absolute value the first eigenvalue of $\tilde H$, which in this case is equal to 1. According to Eq. (\ref{eqn:z2}), this behavior yields a non-trivial topological insulator.
From Re$\Sigma(\omega\!\rightarrow\!\infty)$ we would instead predict a topologically trivial phase. This discovery underlines the crucial role of dynamical fluctuations in the phase-diagram.
Indeed, the high frequency limit of the self energy represents the static part since dynamical fluctuations decay for large absolute values of the frequency and only the Hartree diagrams contribute to the $\lvert \omega\rvert\rightarrow \infty$ limit.
The nontrivial QSH phase survives till Re $\Sigma(\omega\!\rightarrow\!0)$ becomes eventually larger than the largest eigenvalues of $\tilde H$ (in this case equal to 5). This is the case in the Mott insulating state, i.e. after the occupation reaches the (1,1) configuration (slightly above $U\!=\!8$ according to the red curves in Fig. \ref{fig:occ}).
Starting from a topologically trivial band-insulator, the system is therefore driven by $U$ and $J$ first into a non-trivial QSH phase and finally into a trivial Mott insulating state. We note that the topological Mott insulator phase predicted in Ref. \onlinecite{TMI} is qualitatively different from the Mott phase we find in that it describes a topological phase which occurs at strong interactions but can be understood at static mean field level where the Mott transition we observe at strong interactions could not take place.\\

\section{Conclusion}
We studied the topological properties of the BHZ two-band Hubbard model with intra- and inter-band repulsion and a Hund coupling that makes occupation of different orbitals with the same spin energetically favorable. We find that strong interaction eventually favors the Mott state with equal orbital occupation. However, a system with fixed trivial non-interacting band-parameters can go through two phase transitions the first of which drives the system into a nontrivial QSH phase and the second of which results in a trivial Mott insulating state.
The topological phase transition into the QSH band insulating state was shown to be driven by dynamical quantum fluctuations if the non-interacting band parameters are chosen deep enough inside the trivial region.
Our results stimulate the search for topological insulators where multi-orbital and interaction effects play a crucial role. Candidate materials for the physics proposed here might be oxide-heterostructures grown in the [111] direction, whose band structure is known to display a Dirac point \cite{XiaoNComm}. Our two $d$-orbital model at half-filling would be relevant for $d^8$ systems such as LaCuO$_3$, LaAgO$_3$ or LaAuO$_3$ grown on substrates able (e.g. by strain) to lift the degeneracy between the $d_{3z^2-r^2}$ and $d_{x^2-y^2}$ orbitals. For such systems indeed the interaction between the two very localized $e_g$ electrons will be perfectly modeled by our $H_{int}$ with a sizable Hund coupling \cite{GeorgesHundReview}.

\section*{Acknowledgements}
We are grateful to N. Parragh, the main developer of the continuous-time quantum Monte Carlo code we have used for all the calculations done here. We also acknowledge intersting discussions with F.F. Assaad, S. Okamoto, R. Thomale, P. Thunstr\"om, and A. Toschi. Furthermore, we would like to thank the Swedish Science Research Council (JCB) as well as the Deutsche Forschungsgemeinschaft, the European Science Foundation, and the Helmholtz Foundation for financial support. GS acknowledges support by the Deutsche Forschungsgemeinschaft (FOR 1162).


\begin{thebibliography}{47}
\bibitem[{not()}]{note}
\bibinfo{note}{{What is called spin in $H_{\text{BHZ}}$ does not necessarily
  refer to real spin eigenstates but instead to Kramers partners. This could be
  real spin eigenstates or states of opposite total angular momentum
  projections.}}
\end{thebibliography}
\end{document}